\documentclass[runningheads]{llncs}
\usepackage[T1]{fontenc}
\def\Snospace~{\S{}}
\usepackage{nicefrac}
\usepackage{siunitx}
\usepackage{listings}
\usepackage{array,framed}
\usepackage{
  color,
  float,
  epsfig,
  wrapfig,
  graphics,
  graphicx
}
\usepackage{textcomp,amssymb}
\usepackage{setspace}
\usepackage{amsfonts}
\usepackage{enumerate}
\usepackage{enumitem}
\usepackage[compatible]{algpseudocode}
\usepackage{graphics}
\usepackage{subfig}
\usepackage{xparse} 
\usepackage{xspace}
\usepackage{multirow}
\usepackage{hyperref}
\usepackage{xfrac}
\usepackage{tabularx}
\usepackage{flushend}
\usepackage{mathptmx,avant}
 \usepackage{
  tikz,
  pgfplots,
  pgfplotstable
}
\usepackage{hyperref}

\usetikzlibrary{
  shapes.geometric,
  arrows,
  external,
  pgfplots.groupplots,
  matrix
}
\usepackage{amsmath}

\pgfplotsset{compat=1.9}

\DeclareGraphicsExtensions{
    .png,.PNG,
    .pdf,.PDF,
    .jpg,.mps,.jpeg,.jbig2,.jb2,.JPG,.JPEG,.JBIG2,.JB2}

\usepackage{Settings/my_commands}
\usepackage{Settings/listing_algorithm}
\usepackage{caption}

\usepackage{svg}
\usepackage{soul}
\setstcolor{red}
\usepackage{colortbl}
\usepackage{multicol}
\usepackage{arydshln}
\usepackage{verbatim}
\usepackage{fancyvrb}
\usepackage{fancyhdr}

\newcommand{\davide}[1]{}
\usepackage{xparse}
\newcommand{\bnm}{\begin{newmath}}
\newcommand{\enm}{\end{newmath}}

\newcommand{\bea}{\begin{eqnarray*}}%
\newcommand{\eea}{\end{eqnarray*}}%

\newcommand{\bne}{\begin{newequation}}
\newcommand{\ene}{\end{newequation}}

\newcommand{\bal}{\begin{newalign}}
\newcommand{\eal}{\end{newalign}}

\newenvironment{newalign}{\begin{align}%
\setlength{\abovedisplayskip}{4pt}%
\setlength{\belowdisplayskip}{4pt}%
\setlength{\abovedisplayshortskip}{6pt}%
\setlength{\belowdisplayshortskip}{6pt} }{\end{align}}

\newenvironment{newmath}{\begin{displaymath}%
\setlength{\abovedisplayskip}{4pt}%
\setlength{\belowdisplayskip}{4pt}%
\setlength{\abovedisplayshortskip}{6pt}%
\setlength{\belowdisplayshortskip}{6pt} }{\end{displaymath}}

\newenvironment{newequation}{\begin{equation}%
\setlength{\abovedisplayskip}{4pt}%
\setlength{\belowdisplayskip}{4pt}%
\setlength{\abovedisplayshortskip}{6pt}%
\setlength{\belowdisplayshortskip}{6pt} }{\end{equation}}

\newcounter{ctr}

\newcounter{mytable}
\def\mytable{\begin{centering}\refstepcounter{mytable}}
\def\endmytable{\end{centering}}

\newcounter{myfig}
\def\myfig{\begin{centering}\refstepcounter{myfig}}
\def\endmyfig{\end{centering}}

\newlength{\saveparindent}
\newlength{\saveparskip}
\newcommand{\E}{{\rm I\kern-.3em E}}

\renewcommand{\eqref}[1]{\mbox{Equation~(\ref{#1})}}

\def \part {part}

\renewcommand{\paragraph}[1]{\vspace*{6pt}\noindent\textbf{#1}\;}

\def \blackslug{\hbox{\hskip 1pt \vrule width 4pt height 8pt
    depth 1.5pt \hskip 1pt}}
\def \qed{\quad\blackslug\lower 8.5pt\null\par}

\newcounter{mynote}[section]

\newcommand\ignore[1]{}

\newcounter{rcnote}[section]

\newcounter{mrnote}[section]

\newcounter{fknote}[section]

\newcounter{anote}[section]

\DeclareMathSymbol{\mlq}{\mathord}{operators}{``}
\DeclareMathSymbol{\mrq}{\mathord}{operators}{`'}

\newcommand{\rhf}[2]{R_{f, \gamma}}

\DeclareDocumentCommand{\edist}{o o}{
  \ensuremath{
    \IfNoValueTF{#1}{{d}}{{\sf d}(#1,#2)}
  }
}

\newcommand{\olrk}[1]{\ifx\nursymbol#1\else\!\!\mskip4.5mu plus 0.5mu\left(\mskip0.5mu plus0.5mu #1\mskip1.5mu plus0.5mu \right)\fi}

\NewDocumentCommand{\indseq}{ O{1} O{r} }{{#1}\ldots {#2}}

\usepackage{Settings/my_commands}
\usepackage{Settings/listing_algorithm}
\usepackage{caption}
\begin{document}

\title{Unveiling ECC Vulnerabilities: LSTM Networks for Operation Recognition in Side-Channel Attacks}

\author{Alberto Battistello\inst{1} \and Guido Bertoni\inst{1} \and Michele Corrias\inst{1} \and Lorenzo Nava\inst{1} \and Davide Rusconi\inst{2} \and Matteo Zoia\inst{2} \and Fabio Pierazzi\inst{3} \and Andrea Lanzi\inst{2}}

\authorrunning{Battistello et al.}

\institute{Security Pattern, Milan, Italy \and
University of Milan, Milan, Italy \and
King's College London, London, United Kingdom}

\maketitle 

\begin{abstract}
We propose a novel approach for performing side-channel attacks on elliptic curve cryptography. Unlike previous approaches and inspired by the ``activity detection'' literature, we adopt a long-short-term memory (LSTM) neural network to analyze a power trace and identify patterns of operation in the scalar multiplication algorithm performed during an ECDSA signature,  that allows us to recover bits of the ephemeral key, and thus retrieve the signer's private key. Our approach is based on the fact that modular reductions are conditionally performed by micro-ecc and depend on key bits. 

We evaluated the feasibility and reproducibility of our attack through experiments in both simulated and real implementations. We demonstrate the effectiveness of our attack by implementing it on a real target device, an STM32F415 with the micro-ecc library, and successfully compromise it. 
Furthermore, we show that current countermeasures, specifically the coordinate randomization technique, are not sufficient to protect against side channels. Finally, we suggest other approaches that may be implemented to thwart our attack. 

\keywords{Hardware security \and Side-channel attacks \and Elliptic curve cryptography \and Key recovery \and Deep learning}
\end{abstract}

\section{Introduction}

Within the security domain, Machine Learning (ML) methods have been utilized to address issues such as email spam filtering \cite{tretyakov2004machine,crawford2015survey}, intrusion detection systems \cite{tsai2009intrusion,buczak2015survey}, and facial recognition~\cite{taigman2014deepface,oravec2014feature}. Although these uses highlight the effectiveness of ML in protective security strategies, its influence also spans offensive actions like side-channel attacks (SCAs). In SCAs, attackers examine physical attributes of devices, such as timing variations, power usage, and electromagnetic signals during computations, to uncover confidential information, emphasizing the dual use nature of ML in the security field \cite{kocher1996timing,kocher1999differential,quisquater2001electromagnetic}.

This duality is particularly prominent in cryptographic applications, such as the Elliptic Curve Digital Signature Algorithm (ECDSA), where physical attributes can reveal critical secret information, presenting a significant vulnerability. Our research focuses on side channel attacks (SCA) against ECDSA implementations, particularly those \textit{ protected with the coordinate randomization mechanism}, as discussed in~\cite{rivain2011fast}. Based on this foundation, our study proposes an innovative technique using a long- and short-term memory (LSTM) network architecture~\cite{yu2019review} to identify the execution patterns of operations in the scalar multiplication process, an essential part of ECDSA. By examining these operational patterns, our method enables the extraction of ephemeral key bits, potentially leading to the compromise of the signer's full private key. Our study explores the real-world implementation of these insights in elliptic curve cryptography, specifically using the micro-ecc library~\cite{microecc}. This is a widely used open source library, especially common in Internet of Things (IoT) applications.

Luo et al.~\cite{luo2018effective} illustrated the possibility of attacking the ECDSA by using collision attacks, which leverage the detection of specific operational patterns in the traces of power consumption. In these attacks, "collision" signifies the recurrence of an identical value at various stages of the algorithm, identifiable through pattern analysis and correlation methods. Our research improves the attack strategy discussed by~\cite{luo2018effective}, presenting a more efficient SCA approach to exploit vulnerabilities in micro-ecc. By employing a machine learning-based approach and lattice techniques, our method concepts the detection of key operations and recovery of the signer's key, even with a few known ephemeral key bits from signature processes. Additionally, we show that our technique does not depend on electromagnetic (EM) analysis by effectively extracting key information from the noisy power consumption traces of advanced hardware such as the ARM Cortex-M4 processor in the STM32F4 series. This discovery represents a considerable advancement in side-channel attack strategies. Our results also question the efficacy of existing countermeasures against collision attacks, prompting us to suggest more robust alternatives that offer better protection against these and similar vulnerabilities. 

More in detail our framework stands out by employing Long-Short-Term Memory (LSTM) networks, which are well regarded for their proficiency in tasks analogous to human activity recognition. In particular, this approach enhances our analysis from basic side-channel attack execution to "operation recognition," akin to the methods used in detecting human activities. Using this strategy, we can analyze patterns in cryptographic operations in a detailed way, which is essential to detect hidden vulnerabilities. The LSTM network was specifically chosen for its aptitude in deciphering the sequential and temporal dynamics inherent in our attack's operational recognition phase. This decision was informed by the nature of our problem, which aligns more with operational recognition, a concept akin to human activity recognition, than with classic data classification. This delineation underlines the unsuitability of Convolutional Neural Networks (CNNs) for our analysis, as they fall short in capturing the essential temporal context. The LSTM's design, known for its proficiency in processing time-dependent data, perfectly aligns with the requirements of identifying and analyzing execution operational patterns, reinforcing our methodological choices with the goals of our side-channel attack strategy.

To evaluate the practicality of our attack technique, we employed the secp160r1 curve in a simulated setup, executing the attack using an LSTM-based neural network to determine the possibility of extracting the signer's private key. We conducted a successful experiment on the STM32F415 microcontroller~\cite{stmicroelectronics}, selected for its incorporation of the micro-ecc library. This library implements co-Z algorithms and includes a \textit{coordinate randomization countermeasure} to defend against known side-channel attacks (SCAs). To obtain the necessary side-channel traces for our investigation, we utilized the ChipWhisperer~\cite{chipwhisperer} platform, which allowed us to collect power consumption data as leakage vector. Using the same LSTM-based Neural Network (NN) model that was effective in our simulated settings, we modified and trained it specifically for this microcontroller. This was a crucial step in demonstrating that our attack strategy, which was refined through simulations and theoretical models, is not only conceptually sound but also practically applicable to actual hardware. This paper offers the following contributions:

\begin{itemize}

\item Developed an LSTM-based methodology for identifying operations execution patterns in scalar multiplication algorithms, enabling the extraction of ephemeral key bits in ECC.

\item Demonstrated the practicality of this approach through real-world testing on the STM32F415 device, utilizing the micro-ecc library and the secp160r1 curve with the \textit{protected ccoordinate randomization mechanism}, highlighting its adaptability to various types of curves.

\item Illustrated the limitations of existing countermeasures against side-channel attacks while providing an in-depth analysis of potential improvements in security protocols.

\end{itemize}

\section{Background}
\label{sec:background}

This Section provides an introduction to the basic cryptographic concepts necessary to understand the rest of this work. The following Section explores the specific cryptosystem solutions for elliptic curves that are implemented in the target library.
Consider an elliptic curve $\curve_\field$ defined over a field $\field$ of characteristic $\neq 2,3$ according to the short Weierstrass equation as:
\begin{equation}
\label{eq:ec_weierstrass}
    \mathcal{E}: y^2 = x^3 + ax + b \quad a, b \in \mathbb{K}
\end{equation}
with $a,b \in \field$ such that $4a^3+27b^2 \neq 0$. Given the curve, the set of points that satisfy its equation $(x, y) \in \mathbb{K}^2$, augmented with a particular point $\infinitePoint$ called \emph{point at infinity}, forms a \emph{additive abelian group} \cite{de1997elliptic}. 

\subsubsection{Scalar Multiplication}
\label{scalar}
A crucial operation on which protocols based on ECC are based is \emph{scalar multiplication} \cite{lopez2000overview}. Given an integer $k \in \field$ and a point $P$ on $\mathcal{E}$, the scalar multiplication $k$ with $P$ is defined as:
\begin{equation}
    \label{eq:scalarMul}
    kP = \underbrace{P + P + \dots + P}_{k \textsc{ times}}
\end{equation}
In this work we will focus on the scalar multiplication algorithm implemented in micro-ecc, the Montgomery Ladder
implementation with Jacobian co-Z coordinates [10], with coordinates randomization.

\subsubsection{Jacobian co-\textit{Z} representation}
\label{coordinates}
The Jacobian co-\textit{Z} representation
is a projective representation of the points of an elliptic curve that uses three coordinates.

In this system the input points of the operations are represented by triplets sharing the same Z coordinate. This representation was first suggested by Meloni~\cite{meloni2007new}, and then adapted with different trade-offs by Rivain~\cite{rivain2011fast}, and implemented in micro-ecc~\cite{microecc}.

\begin{algorithm}[caption={Montgomery ladder with \textit{(X, Y)}-only co-\textit{Z} addition}, label={alg:montgomeryLadderCoZ}]
input: $P \in \curve(\finiteFieldP)$, $k = (k_{n-1},\dots,k_1,k_0)_2 \in \naturals$/+\text{ with }+/$k_{n-1} = 1$
output: $kP$
begin
    $(R_1,R_0) \gets \texttt{XYCZ-IDBL}(P)$
    for $i \gets n - 2$ to $1$ do
        $b \gets k_i$
        $(R_{1-b},R_b) \gets \texttt{XYCZ-ADDC}(R_b, R_{1-b})$
        $(R_b,R_{1-b}) \gets \texttt{XYCZ-ADD}(R_{1-b}, R_b)$
    end
    $b \gets k_0$
    $(R_{1-b},R_b) \gets \texttt{XYCZ-ADDC}(R_b, R_{1-b})$
    $\lambda \gets \texttt{FinalInvZ}(R_0, R_1, P, b)$
    $(R_b,R_{1-b}) \gets \texttt{XYCZ-ADD}(R_{1-b}, R_b)$
    return $(X_0\lambda^2, Y_0\lambda^3)$
end
\end{algorithm}

\subsubsection{Montgomery ladder with \textit{(X, Y)}-only co-\textit{Z} addition}
\label{para:mlcoz}
The \emph{Montgomery ladder with (X, Y)-only co-Z addition}~\cite{rivain2011fast} \autoref{alg:montgomeryLadderCoZ} is an algorithm employing on Jacobian co-\textit{Z} coordinates 
used in the target ECC implementation~\cite{microecc}
to perform the scalar multiplication required for ECDSA. It performs an initial doubling (in micro-ecc it uses the Jacobian Doubling algorithm), then starts a loop on the bits of the scalar, where for each loop cycle a XYcZ-ADDC  followed by a XYcZ-ADD are executed.\\\\

\begin{algorithm}[caption={XYcZ-ADDC}, label={alg:XYCZ-ADDC}]
input: $(X_1,Y_1)$ and $(X_2,Y_2)$ s.t. $P \equiv (X_1 : Y_1 : Z)$ and $Q \equiv (X_2 : Y_2 : Z)$ for some $Z \in \F_q, P,Q \in \mathcal{E}(\F_q)$
output: $(X_3, Y_3)$ and $(X_3',Y_3')$ s.t. $P + Q \equiv (X_3 : Y_3 : Z_3)$ and $P - Q \equiv (X_3' : Y_3' : Z_3)$ for some $Z_3 \in \F_q$
begin
    $A \gets (X_2 - X_1)^2$
    $B \gets X_1A$
    $C \gets X_2A$
    $D \gets (Y_2 - Y_1)^2$
    $F \gets (Y_1 + Y_2)^2$
    $E \gets Y_1(C - B)$
    $X_3 \gets D - (B + C)$
    $Y_3 \gets (Y_2 - Y_1)(B - X_3) - E$
    $X_3' \gets F - (B + C)$
    $Y_3' \gets (Y_1 + Y_2)(X_3'-B) - E$
    return $((X_3,Y_3),(X_3',Y_3'))$
end
\end{algorithm}
\clearpage
\begin{algorithm}[caption={XYcZ-ADD}, label={alg:XYCZ-ADD}]
input: $(X_1,Y_1)$ and $(X_2,Y_2)$ s.t. $P \equiv (X_1 : Y_1 : Z)$ and $Q \equiv (X_2 : Y_2 : Z)$ for some $Z \in \F_q, P,Q \in \mathcal{E}(\F_q)$
output: $(X_3, Y_3)$ and $(X_1',Y_1')$ s.t. $P \equiv (X_1' : Y_1' : Z_3)$ and $P + Q \equiv (X_3 : Y_3 : Z_3)$ for some $Z_3 \in \F_q$
begin
    $A \gets (X_2 - X_1)^2$
    $B \gets X_1A$
    $C \gets X_2A$
    $D \gets (Y_2 - Y_1)^2$
    $E \gets Y_1(C - B)$
    $X_3 \gets D - (B + C)$
    $Y_3 \gets (Y_2 - Y_1)(B - X_3) - E$
    $X_1' \gets B$
    $Y_1' \gets E$
    return $((X_3,Y_3),(X_1',Y_1'))$
end
\end{algorithm}

\section{Related Work}

In the physical security domain, the adoption of Neural Networks (NNs) has marked a transformative phase, particularly in enhancing Side-Channel Attack (SCA) strategies. These NN-empowered SCAs have surpassed traditional approaches by yielding more potent results with reduced observational demands \cite{maghrebi2016breaking,picek2017side,wang2014learning,wu2021best,perin2021keep,nascimento2017applying,weissbart2019one,cagli2017convolutional,zaid2020methodology,carbone2019deep,picek2021sok}. Research initiatives \cite{maghrebi2019deep,ramezanpour2020scaul,benadjila2020deep}, have taken on sophisticated Deep Learning techniques to exploit side channel traces in examining symmetric algorithms. Focusing on the ECC Double-And-Add-Always algorithm implemented on FPGA platforms, Mukhtar et al.\cite{mukhtar2018machine} applied classification methods to reveal secret key bits of the ECC. In a similar vein, Weissbart et al.\cite{weissbart2019one} orchestrated a power analysis attack on the Edwards-curve Digital Signature Algorithm (EdDSA)\cite{bernstein2012high}, revealing the superior capabilities of CNN over classical side-channel techniques like Template Attacks~\cite{chari2002template}. Weissbart et al.\cite{weissbart2020systematic} further expanded their investigation, evaluating additional protected targets and highlighting the efficacy of Deep Learning, particularly CNNs, in breaching protected implementations of scalar multiplication on Curve25519\cite{bernstein2006curve25519}.

Perin et al.\cite{perin2021keep} proposed a groundbreaking Deep Learning-based iterative framework for unsupervised horizontal attacks, aimed at refining the accuracy of single-trace attacks and reducing errors in the decryption of private keys, particularly in protected ECC implementations. This effort, similar to Nascimento et al.\cite{nascimento2017applying}, exploited vulnerabilities within the $\mu$NaCl library's \texttt{cswap} function~\cite{NACLlib}, showcasing the ongoing evolution of NN-enabled SCA methodologies in enhancing cryptographic security. In a recent development, Staib et al.\cite{staib2023deep} sought to advance collision side channel attacks through deep learning, demonstrating a neural network's superiority in collision detection over traditional methods on a public dataset\cite{luo2018effective,clavier2011improved,bauer2015horizontal}. 

Our research uses long- and short-term memory (LSTM) networks, which are renowned for their effectiveness in tasks that resemble recognition of human activity. This innovative approach not only goes beyond the traditional scope of side-channel attacks (SCA) but reframes the problem as an "operation recognition" task. By adopting methodologies similar to those in human activity detection, we enable sophisticated pattern recognition within cryptographic operations. This unique framework is pivotal for uncovering hidden vulnerabilities, highlighting the intricate interplay between cryptographic processes and exploitable weaknesses.

\section{Attack Preconditions}
\label{sec:attackz}

Our objective for this attack is the \textit{micro-ecc} library, which is an open-source implementation of ECDH and ECDSA tailored for 8/32/64 bit processors \cite{microecc}. Although this library supports multiple elliptic curves, our study specifically targets \textit{secp160r1}. This choice is driven by the widespread use of the curve in resource-limited environments such as IoT, where both security and efficiency are paramount, making it an important focus of our investigation. Our target utilizes the Jacobian Co-\textit{Z} representation along with the Montgomery ladder algorithm (\autoref{alg:montgomeryLadderCoZ}), for cryptographic processes. Furthermore, the library incorporates a countermeasure of coordinate randomization, which produces a random $z \in \finiteFieldQ$ before each execution. This random $z$ is then applied to the representation of points before entering the Montgomery ladder. This initial randomization ensures that the values computed during the ladder are not correlated with the original point, making modular reductions unpredictable based on guesses of bits from the ephemeral key. However, we have discovered several issues that make this implementation susceptible to side-channel attacks, which will be discussed in the following.

\paragraph{Consistent Timing of the Implementation.}
\label{subsec:ct}
The initial problem arises because the micro-ecc code performs a conditional \emph{modular reduction}, which is triggered by an over / overflow occurring after an addition or subtraction operation. This implies that the algorithm does not run in constant time, thereby leaking information during execution. However, exploiting this vulnerability (e.g., intercepting the over/underflow operations) is made difficult by the randomization of the coordinates. This countermeasure effectively nullifies any predictive attempts without prior knowledge of the randomness used in the countermeasure, thus preventing exploitation of the issue \textit{ as is}.

\paragraph{Repeated Operations Depending on Key Bit.}The second vulnerability in the micro-ecc implementation arises from the Montgomery ladder (\autoref{alg:montgomeryLadderCoZ}) performing certain calculations twice, based on the ephemeral key bit. Specifically, during the $(n-1)$\textsuperscript{th} iteration of the Montgomery ladder, the value $B - X_3$ is calculated. Subsequently, both $B$ and $X_3$ are returned as the x-coordinates of the two output points ($X_1'$ and $X_3$, respectively). In the next $n$\textsuperscript{th} iteration of the Montgomery ladder, depending on the $n$\textsuperscript{th} bit of the scalar, the implementation either recalculates $B - X_3$ or computes $X_3 - B$. This means that whenever two consecutive bits of the scalar are identical, the $B - X_3$ operation is performed twice. 

\paragraph{Attacking the First Bit.} The final issue identified in the target system is a leak resulting from the interaction between the Jacobian doubling operation, performed before the Montgomery ladder, and the addition operation executed during it. Specifically, the value of the first key bit determines whether the same subtraction is performed or its inverse is calculated. Consequently, detecting a collision between the subtraction within the Jacobian doubling and the first iteration of the Montgomery ladder reveals the first bit of the ephemeral key.
\\\\
Our attack exploits all three identified issues to leak the initial bits of the ephemeral key and subsequently infer the private key. Specifically, we used a timing side channel to detect whether an overflow occurred during the addition operation, leveraging the first issue. Then, by distinguishing between overflowing and non-overflowing subtractions, we directly leak the first bit using the third issue. Additionally, we exploit the second vulnerability to infer the equality of the first bits. This information enables us to mount a lattice reduction attack, effectively extracting the private key used in the signatures, thereby compromising the system's security. To achieve this, we need to design a neural network capable of identifying the operating patterns based on the power trace.

\section{System Overview}
\label{sec:sys_ov}
To exploit the vulnerability described in Section \ref{sec:attackz}, we have developed a system consisting of five main components: the acquisition unit responsible for acquiring the power traces from the target device, the windowing algorithm is primarily designed to divide the power traces into smaller segments (e.g., windows), a machine learning model that categorizes the acquired traces, a post-processing unit that analyzes the output of the ML model to identify the presence or absence of modular operations within the traces, and a final component that deduces the scalar bits as discussed in Section \ref{sec:attackz} and extracts the key.

\begin{figure*}[h]
    \centering\includegraphics[width=0.95\textwidth]{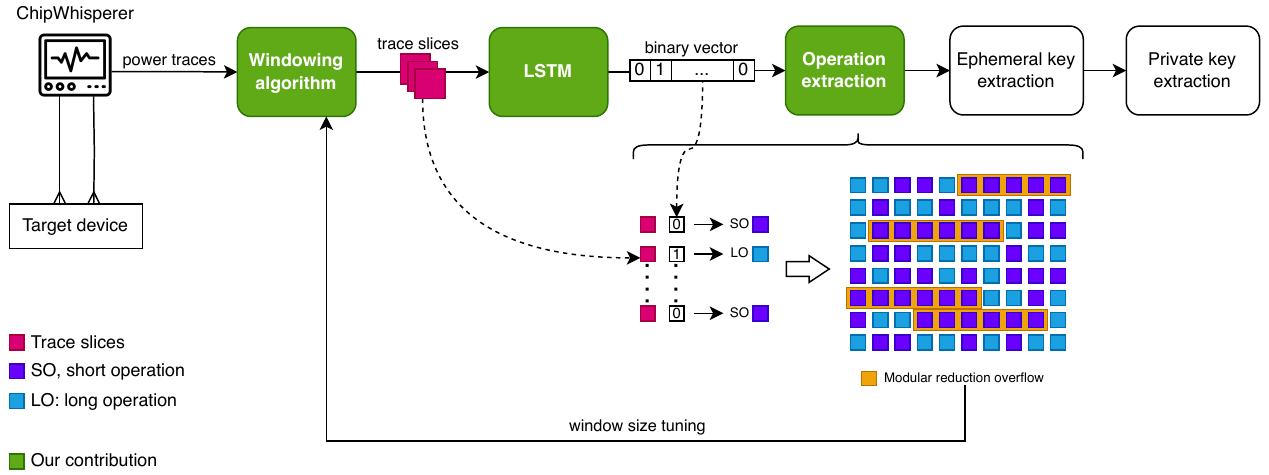}
    \caption{Architectural Overview}  \label{fig:sys_arch}
\end{figure*}

As illustrated in Figure~\ref{fig:sys_arch}, the process begins with recording raw power traces from the targeted device through an acquisition unit. These raw traces are then preprocessed and divided into multiple windows of uniform size (e.g, windowing algorithm). Each window is subsequently fed into the neural network, where our model determines whether the operation within the window is a short operation (SO), such as addition or subtraction, or a long operation (LO), such as multiplication and division. Based on the neural network's classification, the window vector is transformed into a binary vector, enabling us to identify the type of operation executed. For short operations, we also check for overflow. This generates a list of operations and details concerning modular reductions, which can be utilized, following the leaks in \autoref{sec:attackz}, to determine the values and relationships of the bits of the ephemeral key used in the attacked ECDSA round. With sufficient leakage information on the ephemeral key, we can proceed to recover the private key using a known lattice reduction technique, thus compromising the target and completing the attack.

\section{System Architecture}
\label{sec:sys}

We will now describe the design of each component of our framework, except for the unit responsible for obtaining power consumption traces, as its design depends on the target device. Each component represents a part of our contribution and has been designed to target the cryptographic algorithm.

\subsection{Pre-Processing}

As described in \autoref{sec:sys_ov}, we employ a sliding window algorithm to segment the entire trace into smaller, fixed-size sequences. This method offers several advantages for our analysis and is particularly beneficial when used in conjunction with long-short-term memory (LSTM) networks, known for their effectiveness in tasks similar to recognizing human activities.

Firstly, the use of a sliding window enables us to focus on manageable chunks of data, which simplifies the computational process. By tuning the window size and the offset between adjacent windows, both of which are hyperparameters, we can optimize our analysis for both performance and accuracy. After rigorous testing and evaluation, we determined that a window size of 500 samples and an offset of 10 samples provide the best results. The chosen window size of 500 samples produces an optimal balance by ensuring that each segment contains sufficient detail to accurately classify the operations within it, while also maintaining computational efficiency. A larger window might capture more detail but would significantly increase computational overhead, potentially slowing down the analysis. In contrast, a smaller window might be more efficient, but could miss critical information necessary for accurate classification. The 10-sample offset ensures that we achieve a high-resolution analysis by overlapping windows, allowing for a more granular examination of the data. This overlap helps to ensure that no important transitions or details are missed between windows, enhancing the overall accuracy of our classification. Integrating the sliding-window approach with LSTM networks offers significant benefits. LSTMs are designed to capture temporal dependencies and trends within sequential data, which is crucial for accurate classification.

From a machine learning perspective, splitting the data into windows can significantly improve the efficacy of our LSTM model. By creating a larger number of training samples from the original dataset, we expose the LSTM model to a greater variety of patterns and transitions within the data. This increased dataset size allows the model to learn more effectively and generalize better. Moreover, by tuning the window size and offset, we can control the granularity of the input data, allowing the model to focus on the most relevant features and patterns. 

\subsection{Neural Network Architecture \& Parameters}\label{sec:nn}

In this study, we created a neural network framework aimed at effectively handling binary classification tasks, integrating both convolutional and long-short-term memory (LSTM) layers. This framework is intended to recognize intricate patterns and temporal relationships within the data, thus enhancing the model's predictive accuracy. Our framework is designed to balance computational efficiency with the capacity to extract and utilize significant features, leading to a reliable and robust model. The neural network architecture selected for our framework is a sequential model consisting of the following layers.

\begin{itemize}[noitemsep]
\item \textbf{Convolutional Layer:} This layer performs convolution operations on the input data to decrease its dimensions and transform its structure, thus increasing computational efficiency. The convolutional layers enhance the resilience to noise and scale changes in the input traces. In particular, this layer contains 64 feature maps, each with a 3x3 kernel, and applies the ReLU activation function to incorporate nonlinearity, allowing the network to capture intricate patterns.

\item \textbf{Pooling Layer:} Following the convolutional layer, this layer performs average pooling, reducing the dimensionality of the feature maps produced by the convolutional operations. The pooling layer uses a pooling size of 10, which helps in further down-sampling the input and retaining the most salient features.

\item \textbf{LSTM Layer:} This layer forms the core of our neural network and consists of 1000 internal units. LSTM (Long Short-Term Memory) networks are adept at capturing long-term dependencies and temporal correlations within sequential data. Using previous sample information, the LSTM layer is able to make context-sensitive evaluations, crucial for tasks involving temporal sequences.

\item \textbf{Dropout Layer:} To prevent overfitting and enhance the generalization capabilities of the network, this layer randomly drops a subset of its nodes during training. The dropout rate is set to 0.5, which means that 50\% of the nodes are temporarily ignored during each training iteration. This technique helps the network develop more robust features by mitigating the dependence on any specific subset of neurons.

\item \textbf{Dense Layer:} This fully connected layer comprises 1000 neurons and employs the ReLU activation function. The dense layer integrates information from the preceding layers and contributes to the network's ability to perform high-level abstractions and complex decision making.

\item \textbf{Output Layer:} The final layer of the network is responsible for the binary classification task. It consists of dense nodes that utilize the softmax activation function to produce probabilities for the binary classes, enabling the network to produce the final classification decision.

\end{itemize}

The LSTM network utilizes the binary cross-entropy loss function for training, with optimization done through stochastic gradient descent (SGD). Inspired by real-world applications in the literature \cite{zhang2022multi}, we chose a classic architecture that is typically used in recognition of human activity. In total, the network comprises 5,263,258 trainable parameters, allocated as follows: \textit{Convolutional Layer:} 256 parameters \textit{LSTM Layer:} 4,260,000 parameters (the main component of our model) \textit{Dense Layer:} 1,001,000 parameters \textit{Output Layer:} 2,002 parameters.

\subsection{Post-Processing to Identify Operations}

In order to transform the model output into a series of operations, a two-step method is utilized. The initial step involves aligning the classified operations within the algorithm. The model output already identifies whether the operations within the windows are short operations (SO) or not, setting the groundwork for further analysis. The second step differentiates between short operations with overflows and those without, which is critical for our attack. This differentiation is based on the observation that overflows cause a modular reduction, leading to an increased duration. Thus, by examining a continuous sequence of windows identified as short operations, we can estimate the duration of each individual operation in the sequence by analyzing the starting points of SOs and the subsequent operations.  Specifically, the procedure involves the following steps:

\begin{enumerate}
    \item \textbf{Classification Matching:} Use the output of the model to determine the particular operations that occur within the algorithm, depending on whether each window is classified as containing a short operation or not.
    \item \textbf{Duration Analysis:} For groups of consecutive windows marked as short operations, determine the estimated duration of each operation. An extended duration can suggest the occurrence of a modular reduction, which indicates an overflow. This distinction uses the pre-determined window length and step size to precisely measure operation durations.
\end{enumerate}

It is crucial to understand that the time taken for a subtraction operation involving modular reduction can differ between various target devices. Therefore, when implementing this post-processing step on a new target device, modifications must be made to consider the unique physical properties of that device. This adaptation guarantees a precise differentiation between operations in diverse hardware contexts. This dual-phase approach improves the ability to correctly detect and distinguish between brief operations with and without overflows, thereby enhancing the attack's overall efficiency.

\subsection{Collision Template \& Leaking the Ephemeral Key}
Following the identification operations process, it is necessary to deploy a collision attack template to retrieve the bits of the ephemeral key. This template can indicate the locations where collisions occur in the algorithm's operations, and from these collisions one can infer the positions and values of the ephemeral key bits. This phase relies on the specific implementation of the cipher algorithm and must be adapted if the algorithm is modified. To construct the collision template model and extract the ephemeral key bits, we started with a simulation of the target algorithm, as outlined in Section \ref{sec:eval}. This simulation gives us insight into the intermediate values during the computation and the sequence of executed instructions, noting the occurrences of overflows and underflows. Using these insights and the knowledge from \autoref{sec:attackz}, we identified specific operations that reveal ephemeral key bits, even with coordinate randomization countermeasures in effect.

In~\autoref{tab:key_op01}, we present examples of the algorithm simulation that selected operations across different initial bits and multiple keys within the coordinate randomization defense framework. These operations include the Initial Double and the first iteration of the Montgomery loop, with ``\texttt{||}'' indicating the start of the loop. The operations are labeled as ``\texttt{A}'' for ``\emph{Add}, \texttt{S}'' for \emph{Sub}, and ``\texttt{M}'' for both \emph{Mul} and \emph{Sqr}, with a ``\texttt{+}'' next to an operation representing a modular reduction. 
A \emph{collision} is described as the occurrence of identical underflow behavior in two operations, underscoring their importance in our analysis.

\begin{table}[]
\small
\caption{Operations for the \texttt{initial\_double} and the first iteration of the Montgomery loop, for values of the 3 initial bits of the ephemeral key. Randomization countermeasure is active (i.e., \emph{z} is random). In red are highlighted operation 18 and 26, and the bit of the key that correlates with their collision.}\label{tab:key_op01}
\centering
\scalebox{0.7}{
\resizebox{\columnwidth}{!}{%
\begin{tabular}{r|l}
\emph{Key bits} & \emph{Operations} \\
\hline
position & \texttt{0 1 2 3 4 5 6 7 8 9 10\! 11\! 12\!\! 13\!\! 14\! 15\!\! 16\!\! 17\!\! {\color{red}18}\!\! 19\!\! 20\!\! 21\!\! 22\!\! 23\!\!\! 24\!\! 25\!\!\! {\color{red}26}\!\! 27\!\! 28\!\! 29 } \\
0b1{\color{red}1}0 & \texttt{M  M  M  M  M  M  M  M  M A }\space \texttt{A+}       \texttt{S }\space \texttt{M A }\space \texttt{A+}       \texttt{M S+}       \texttt{S+}       \texttt{\textcolor{red}{S+}}       \texttt{M S }\space \texttt{M  M  M  M} \texttt{||} \texttt{\textcolor{red}{S+}}       \texttt{M  M  M} \\
0b1{\color{red}0}0 & \texttt{M  M  M  M  M  M  M  M  M A }\space \texttt{A+}       \texttt{S }\space \texttt{M A+}       \texttt{A+}       \texttt{M S+}       \texttt{S }\space \texttt{\textcolor{red}{S }}\space \texttt{M S+}       \texttt{M  M  M  M} \texttt{||} \texttt{\textcolor{red}{S+}}       \texttt{M  M  M} \\
0b1{\color{red}0}0 & \texttt{M  M  M  M  M  M  M  M  M A+}       \texttt{A+}       \texttt{S+}       \texttt{M A }\space \texttt{A+}       \texttt{M S }\space \texttt{S+}       \texttt{\textcolor{red}{S }}\space \texttt{M S+}       \texttt{M  M  M  M} \texttt{||} \texttt{\textcolor{red}{S+}}       \texttt{M  M  M} \\
0b1{\color{red}0}0 & \texttt{M  M  M  M  M  M  M  M  M A }\space \texttt{A }\space \texttt{S }\space \texttt{M A }\space \texttt{A }\space \texttt{M S }\space \texttt{S }\space \texttt{\textcolor{red}{S+}}       \texttt{M S+}       \texttt{M  M  M  M} \texttt{||} \texttt{\textcolor{red}{S }}\space \texttt{M  M  M} \\
0b1{\color{red}0}1 & \texttt{M  M  M  M  M  M  M  M  M A+}       \texttt{A+}       \texttt{S+}       \texttt{M A }\space \texttt{A+}       \texttt{M S }\space \texttt{S+}       \texttt{\textcolor{red}{S+}}       \texttt{M S }\space \texttt{M  M  M  M} \texttt{||} \texttt{\textcolor{red}{S }}\space \texttt{M  M  M} \\
0b1{\color{red}1}1 & \texttt{M  M  M  M  M  M  M  M  M A }\space \texttt{A }\space \texttt{S+}       \texttt{M A+}       \texttt{A+}       \texttt{M S }\space \texttt{S+}       \texttt{\textcolor{red}{S }}\space \texttt{M S+}       \texttt{M  M  M  M} \texttt{||} \texttt{\textcolor{red}{S }}\space \texttt{M  M  M} \\
0b1{\color{red}0}1 & \texttt{M  M  M  M  M  M  M  M  M A+}       \texttt{A+}       \texttt{S+}       \texttt{M A }\space \texttt{A+}       \texttt{M S }\space \texttt{S+}       \texttt{\textcolor{red}{S }}\space \texttt{M S }\space \texttt{M  M  M  M} \texttt{||} \texttt{\textcolor{red}{S+}}       \texttt{M  M  M} \\
0b1{\color{red}0}1 & \texttt{M  M  M  M  M  M  M  M  M A }\space \texttt{A+}       \texttt{S }\space \texttt{M A }\space \texttt{A }\space \texttt{M S+}       \texttt{S }\space \texttt{\textcolor{red}{S }}\space \texttt{M S+}       \texttt{M  M  M  M} \texttt{||} \texttt{\textcolor{red}{S+}}       \texttt{M  M  M} \\
0b1{\color{red}1}0 & \texttt{M  M  M  M  M  M  M  M  M A+}       \texttt{A+}       \texttt{S }\space \texttt{M A+}       \texttt{A }\space \texttt{M S+}       \texttt{S+}       \texttt{\textcolor{red}{S }}\space \texttt{M S+}       \texttt{M  M  M  M} \texttt{||} \texttt{\textcolor{red}{S }}\space \texttt{M  M  M} \\
\end{tabular}
}
}
\end{table}

\begin{table}[!ht]
\caption{Operations for the first iteration of the Montgomery loop, for random values of the 3 initial bits of the ephemeral key.}\label{tab:key_op55_59}
\centering
\scalebox{0.98}{
\resizebox{\columnwidth}{!}{%
\begin{tabular}{r|l}
\emph{Key bits} & \emph{Operations} \\
\hline
position & \texttt{24\!\! 25\!\! 26\!\! 27\!\! 28\!\! 29\!\! 30\!\! 31\!\! 32\!\! 33\!\! 34\!\! 35\!\! 36\!\! 37\!\! 38\! 39\!\! 40\!\! 41\!\! 42\! 43\!\! 44\!\! 45\!\! 46\!\! 47\!\! 48\!\! 49\!\! 50\!\! 51\!\! 52\!\! 53\!\! 54 {\color{red}55} 56\! 57\! 58\! {\color{red}59}\! 60\!\! 61\!\! 62\!\! 63\!\!} \\
 0b1{\color{red}10} & \texttt{M}\space \texttt{||}\space \texttt{*S+}\space \texttt{M  M  M A }\space \texttt{S }\space \texttt{S+}\space \texttt{M A+}\space \texttt{M S }\space \texttt{S }\space \texttt{M S+}\space \texttt{M S }\space \texttt{S+}\space \texttt{M S }\space  \texttt{S+}\space \texttt{M  M  M S+}\space \texttt{M S }\space \texttt{S+}\space \texttt{S }\space \texttt{M \textcolor{red}{*S+}}\space \texttt{M S+}\space \texttt{||}\space \texttt{\textcolor{red}{*S }}\space \texttt{M M M A+} \\
 0b1{\color{red}00} & \texttt{M}\space \texttt{||}\space \texttt{*S+}\space \texttt{M  M  M A+}\space \texttt{S+}\space \texttt{S+}\space \texttt{M A }\space \texttt{M S+}\space \texttt{S }\space \texttt{M S+}\space \texttt{M S+}\space \texttt{S }\space \texttt{M  S+}\space \texttt{S }\space \texttt{M  M  M S+}\space \texttt{M S }\space \texttt{S+}\space \texttt{S }\space \texttt{M \textcolor{red}{*S+}}\space \texttt{M S }\space \texttt{||}\space \texttt{\textcolor{red}{*S+}}\space \texttt{M M M A } \\
 0b1{\color{red}00} & \texttt{M}\space \texttt{||}\space \texttt{*S+}\space \texttt{M  M  M A+}\space \texttt{S }\space \texttt{S }\space \texttt{M A }\space \texttt{M S+}\space \texttt{S }\space \texttt{M S }\space \texttt{M S+}\space \texttt{S }\space \texttt{M  S+}\space \texttt{S }\space \texttt{M  M  M S }\space \texttt{M S }\space \texttt{S+}\space \texttt{S }\space \texttt{M \textcolor{red}{*S+}}\space \texttt{M S+}\space \texttt{||}\space \texttt{\textcolor{red}{*S+}}\space \texttt{M M M A+} \\
 0b1{\color{red}00} & \texttt{M}\space \texttt{||}\space \texttt{*S\space}  \texttt{M  M  M A+}\space \texttt{S }\space \texttt{S }\space \texttt{M A+}\space \texttt{M S+}\space \texttt{S+}\space \texttt{M S+}\space \texttt{M S }\space \texttt{S }\space \texttt{M  S+}\space \texttt{S+}\space \texttt{M  M  M S+}\space \texttt{M S+}\space \texttt{S }\space \texttt{S+}\space \texttt{M \textcolor{red}{*S }}\space \texttt{M S+}\space \texttt{||}\space \texttt{\textcolor{red}{*S }}\space \texttt{M M M A+} \\
 0b1{\color{red}01} & \texttt{M}\space \texttt{||}\space \texttt{*S\space}  \texttt{M  M  M A }\space \texttt{S+}\space \texttt{S+}\space \texttt{M A+}\space \texttt{M S }\space \texttt{S }\space \texttt{M S }\space \texttt{M S+}\space \texttt{S }\space \texttt{M  S+}\space \texttt{S }\space \texttt{M  M  M S }\space \texttt{M S }\space \texttt{S+}\space \texttt{S }\space \texttt{M \textcolor{red}{*S }}\space \texttt{M S }\space \texttt{||}\space \texttt{\textcolor{red}{*S+}}\space \texttt{M M M A } \\
 0b1{\color{red}11} & \texttt{M}\space \texttt{||}\space \texttt{*S\space}  \texttt{M  M  M A+}\space \texttt{S }\space \texttt{S }\space \texttt{M A }\space \texttt{M S+}\space \texttt{S }\space \texttt{M S }\space \texttt{M S+}\space \texttt{S }\space \texttt{M S+}\space  \texttt{S }\space \texttt{M  M  M S }\space \texttt{M S }\space \texttt{S+}\space \texttt{S+}\space \texttt{M \textcolor{red}{*S+}}\space \texttt{M S+}\space \texttt{||}\space \texttt{\textcolor{red}{*S+}}\space \texttt{M M M A+} \\
 0b1{\color{red}01} & \texttt{M}\space \texttt{||}\space \texttt{*S+}\space \texttt{M  M  M A }\space \texttt{S+}\space \texttt{S+}\space \texttt{M A }\space \texttt{M S+}\space \texttt{S+}\space \texttt{M S+}\space \texttt{M S+}\space \texttt{S }\space \texttt{M  S+}\space \texttt{S+}\space \texttt{M  M  M S+}\space \texttt{M S }\space \texttt{S+}\space \texttt{S+}\space \texttt{M \textcolor{red}{*S }}\space \texttt{M S+}\space \texttt{||}\space \texttt{\textcolor{red}{*S+}}\space \texttt{M M M A+} \\
 0b1{\color{red}01} & \texttt{M}\space \texttt{||}\space \texttt{*S+}\space \texttt{M  M  M A+}\space \texttt{S }\space \texttt{S }\space \texttt{M A+}\space \texttt{M S+}\space \texttt{S }\space \texttt{M S }\space \texttt{M S+}\space \texttt{S }\space \texttt{M  S+}\space \texttt{S }\space \texttt{M  M  M S }\space \texttt{M S+}\space \texttt{S }\space \texttt{S+}\space \texttt{M \textcolor{red}{*S }}\space \texttt{M S }\space \texttt{||}\space \texttt{\textcolor{red}{*S+}}\space \texttt{M M M A } \\
 0b1{\color{red}10} & \texttt{M}\space \texttt{||}\space \texttt{*S\space}  \texttt{M  M  M A+}\space \texttt{S }\space \texttt{S+}\space \texttt{M A+}\space \texttt{M S }\space \texttt{S }\space \texttt{M S }\space \texttt{M S }\space \texttt{S+}\space \texttt{M S+}\space  \texttt{S }\space \texttt{M  M  M S }\space \texttt{M S }\space \texttt{S+}\space \texttt{S }\space \texttt{M \textcolor{red}{*S+}}\space \texttt{M S }\space \texttt{||}\space \texttt{\textcolor{red}{*S }}\space \texttt{M M M A } \\
\end{tabular}
}
}
\end{table}

In~\autoref{tab:key_op01}, we note that operations 18 and 26 (marked in red) experience \emph{collision} when the second bit of the scalar is ``1'', regardless of the random ``z'' value.
  Similarly, we can deduce the equality of the 2\textsuperscript{nd} and 3\textsuperscript{rd} bits by detecting collisions between operations 55 and 59, as detailed in~\autoref{tab:key_op55_59}. These collisions occur whether both operations undergo a modular reduction as discussed in~\autoref{sec:attackz}. This pattern of identifying bit-equality through operational collisions can be extended iteratively. For example, \autoref{tab:collisions_indexes} lists operation indexes useful for extracting information about the first 6 bits of the ephemeral key. Upon completion of this phase, the component reveals details regarding the collision model as depicted in \autoref{tab:collisions_indexes}, and these indexes will be utilized to identify the bits of the ephemeral key.

\begin{table}[!ht]
    \caption{Correspondence between the index of operations to search for collision (2\textsuperscript{nd} column), the bit of the key retrieved (1\textsuperscript{st} column) and the information provided by a collision (3\textsuperscript{rd} column).}\label{tab:collisions_indexes}
\centering
\scalebox{0.6}{
\resizebox{\columnwidth}{!}{
\begin{tabular}{r|l|l}
\emph{Bit/s} & \emph{Subs indexes} & \emph{Information}\\
\hline
2 & (18,26) &  \mbox{``1'' if collision, ``0'' otherwise}\\
2,3 & (55,59) & \mbox{bits equal if no collision, different otherwise}\\
3,4 & (88,92) & \mbox{bits equal if no collision, different otherwise}\\
4,5 & (121,125) & \mbox{bits equal if no collision, different otherwise}\\
5,6 & (154,158)  &\mbox{bits equal if no collision, different otherwise}\\

\end{tabular}%
}
}
\end{table}

\subsection{Extracting the Private Key}
The final stage in the process involves extracting the private key by performing an LLL attack. Using the earlier-mentioned recognition method, we can obtain segments of the ephemeral key used in ECDSA scalar multiplication. Subsequently, we apply an LLL attack to derive the signer's secret key from the gathered data. Various versions and implementations of the LLL attack are available, and we specifically opted for the FPlll~\cite{fplll} lattice implementation. This choice is motivated by its Python compatibility, support for various LLL reduction techniques, and its accessibility and efficiency. The basis used to represent the extracted values for each ephemeral key in the LLL algorithm aligns with that of~\cite{jancar2020minerva}, followed by Babai's nearest plane reduction. With this setup, we managed to recover the signer's key using as few as 60 signatures, where each signature included an ephemeral key with at least 5 known bits. Additional details on LLL attacks can be found in~\cite{jancar2020minerva}. 

\section{Evaluation Methodology}
\label{sec:eval}

Our evaluation methodology uses a two-phase approach to determine the efficiency of our system. The initial phase utilizes a simulated setting to adjust the neural network (NN) hyperparameters. The subsequent phase assesses the network performance using power traces obtained from a target device, aiming to evaluate the practicality of the proposed attack.

\subsection{Simulated Environment}
The simulated environment aims to improve the neural network's performance on real power consumption data. Therefore, the artificial dataset is designed to accurately represent real-world situations.

To create the synthetic power traces, we started by re-implementing our chosen Montgomery ladder algorithm. We then ran it with varying values and linked each instruction of the algorithm to a sample in the trace. The value of each sample is based on leakage measurements from thousands of previously analyzed real-world traces. This method ensures that the synthetic traces accurately represent the data we can gather in actual scenarios, maintaining coherence in the training and tuning of the model, although it requires retraining when the target device is changed.

To further improve the dataset's realism, we apply a few extra modifications to the traces: to accurately represent modular reductions, we prolong the leakage trace of a subset of the SOs by adding a segment of its base pattern to simulate additional computation time. Additionally, we introduce vertical random noise to each operation type's base leakage pattern, with a mean of 0 and a standard deviation of 1.5. Lastly, we incorporate horizontal random noise to alter the trace's length, which reflects the operational jitter present in the actual target.

Through the aforementioned steps, we obtain a final set of 12 traces. This quantity has been shown to be adequate, as discussed in Section \ref{sec:nn}, because our model processes not the entire traces at once, but the overlapping sliding windows of samples extracted from the power traces. Consequently, this allows us to maintain a manageable trace count while still deriving enough input data from these traces to fully train and test our ML model.
In particular, our dataset comprises 240,800 windows in total, with 160,533 allocated for NN training and 80,267 designated for testing. This represents an approximate 70/30 split of the entire dataset. To ensure that the network does not detect patterns unrelated to the recognition task, we balanced the ratio of SOs and LOs equally across both sets.

\subsection{Real World Target}
\label{sec:exp_real}
\begin{figure}[h]
    \centering\includegraphics[width=\columnwidth]{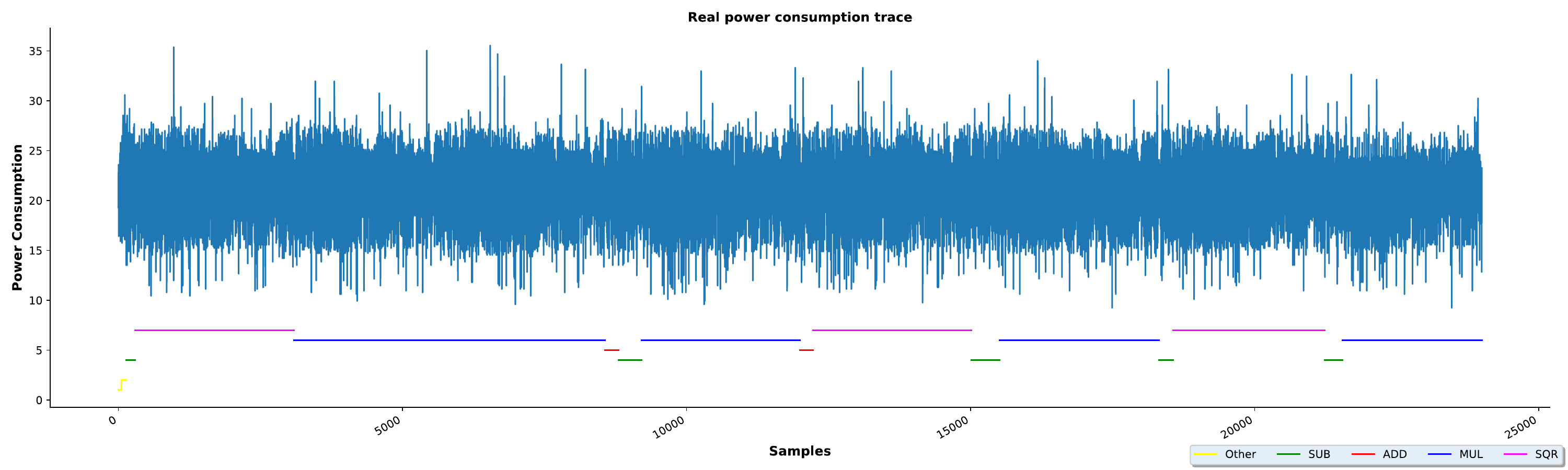}
    \caption{The first 24 thousand samples of a real power consumption trace (translated on the y-axis by 20 points for typesetting reasons).}  \label{fig:real_power_consumption_trace}
\end{figure}

In order to assess our system in a real world scenario, we employed the following configuration. The target device utilized is an STM32F415RG microcontroller integrated within the ChipWhisperer platform. Its core comprises an ARM 32-bit Cortex-M4 CPU operating at speeds of up to 168 MHz. For measuring power consumption, we used the ChipWhisperer acquisition device. This equipment features a 10-bit ADC with a maximum sample rate of 105 MS/s, an AC-Coupled analog input, an adjustable gain up to +55 dB, and it can generate clocks ranging from 5 MHz to 200 MHz. The acquisition of traces was carried out in streaming mode, capturing one sample per clock cycle, at a frequency of 7.37 MHz. This sampling rate was chosen based on the acquisition device's limitations, which supports a maximum of 10 MS/s in streaming mode because of its buffer capacity. To assess the classification results, the test bench also supplied additional metadata for each acquired trace. Specifically, for each segment within the power trace, the operation that produced it is collected, as demonstrated in Figure~\ref{fig:real_power_consumption_trace}.

We obtained a total of five power traces, using four for model training and one for testing. As mentioned earlier, this quantity was more than adequate for training and evaluating the model due to their extensive data. Table \ref{tab:details_realPwerConsumptionTraces} provides a detailed breakdown of these traces, demonstrating that the number of inputs extracted is sufficient for a thorough evaluation of our system.

\begin{table}[h!]
    \centering
    \caption[Real power consumption traces sizes]{Sizes of four real power consumption traces, acquired for testing, with derived datasets. For each power consumption trace (1\textsuperscript{st} column), we provide its samples size (2\textsuperscript{nd} column), the total number of short operation (SO) and long operation (LO) samples (3\textsuperscript{rd} column), the size of the testing dataset (4\textsuperscript{th} column) and the number of \texttt{true}-tagged and \texttt{false}-tagged windows (last column).
    }
    \scalebox{0.6}{
    \label{tab:details_realPwerConsumptionTraces}
    \resizebox{\columnwidth}{!}
    {%
        \begin{tabular}{c||c|cc||cl|cc}

            \textit{Trace} &
            
            \textit{Trace Size} &
            \multicolumn{2}{c||}{
            \begin{tabular}[c]{@{}c@{}}\textit{Trace Samples} \\\textit{\small SO samples} $\;$ \textit{\small LO samples}\end{tabular}} &
            \multicolumn{2}{c|}{
            \textit{Dataset size}} &
            \multicolumn{2}{c}{
            \begin{tabular}[c]{@{}c@{}}\textit{Dataset Windows}\\\texttt{\small true}\textit{\small -tag} $\;$ \texttt{\small false}\textit{\small -tag}\end{tabular}} \\\hline
            
            \texttt{T1} &
            $7,032,251$ &
            \multicolumn{1}{c:}{$\hspace{0.6em} 524,697 \hspace{0.6em}$} &
            $6,507,554$ &
            \multicolumn{2}{c|}{\begin{tabular}[c]{@{}c@{}}$703,176$\end{tabular}} &
            \multicolumn{1}{c:}{$\hspace{0.4em} 86,948 \hspace{0.4em}$} &
            $616,228$ \\ \hline
            
            \texttt{T2} &
            $7,002,209$ &
            \multicolumn{1}{c:}{$\hspace{0.6em} 522,537 \hspace{0.6em}$} &
            $6,479,672$ &
            \multicolumn{2}{c|}{\begin{tabular}[c]{@{}c@{}}$700,171$\end{tabular}} &
            \multicolumn{1}{c:}{$\hspace{0.4em} 86,739 \hspace{0.4em}$} &
            $613,432$ \\ \hline
            
            \texttt{T3} &
            $7,005,521$ &
            \multicolumn{1}{c:}{$\hspace{0.6em} 525,957 \hspace{0.6em}$} &
            $6,479,564$ &
            \multicolumn{2}{c|}{\begin{tabular}[c]{@{}c@{}}$700,503$\end{tabular}} &
            \multicolumn{1}{c:}{$\hspace{0.4em} 87,086 \hspace{0.4em}$} &
            $613,417$ \\ \hline
            
            \texttt{T4} &
            $7,027,186$ &
            \multicolumn{1}{c:}{$\hspace{0.6em} 524,247 \hspace{0.6em}$} &
            $6,502,939$ &
            \multicolumn{2}{c|}{\begin{tabular}[c]{@{}c@{}}$702,669$\end{tabular}} &
            \multicolumn{1}{c:}{$\hspace{0.4em} 86,893 \hspace{0.4em}$} &
            $615,776$ \\
            
        \end{tabular}
    }
    }
\end{table}

\section{Results}
\label{sec:res}
As mentioned in Section \ref{sec:sys}, in order for our system to work, the network must properly classify the windows given in the input, since misclassifying an input can hinder the estimation of the duration of operations.
After training and testing on both datasets detailed in Section \ref{sec:eval} our NN, the core part of our system, proved to be extremely effective both in the simulated environment and, most importantly, in the real world target.
Specifically in the simulated environment, we reached an accuracy of approximately 99.9\%.

\begin{figure}[h!]
\centering
\scalebox{0.6}{
\includegraphics[width=\linewidth]{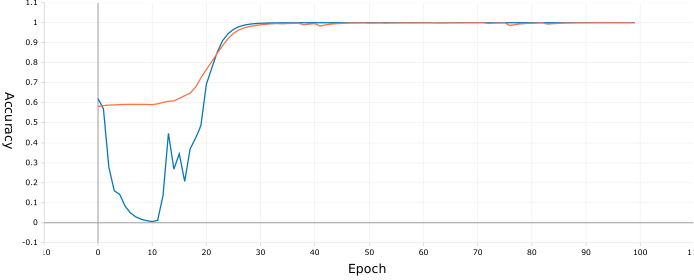}
}
\caption{Epoch accuracy plot for 100 epochs, batch size of 64 and validation split of $20\%$: train in orange, validation in blue.}
\label{fig:epoch_accuracy}
\end{figure}

\begin{figure}[h!]
\centering
\scalebox{0.6}{
\includegraphics[width=\linewidth]{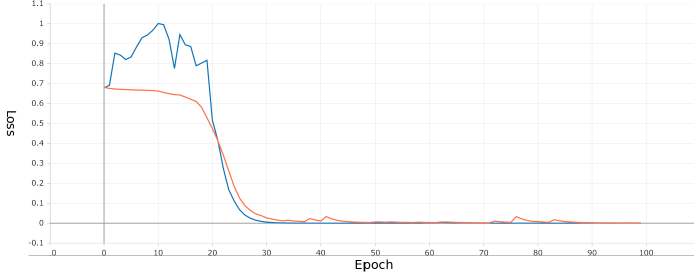}
}
\caption{Epoch loss plot for 100 epochs, batch size of 64 and validation split of $20\%$: train in orange, validation in blue.}
\label{fig:epoch_loss}
\end{figure}
\begin{figure}[h]
    \centering
    \scalebox{0.85}{
    \includegraphics[width=\columnwidth]{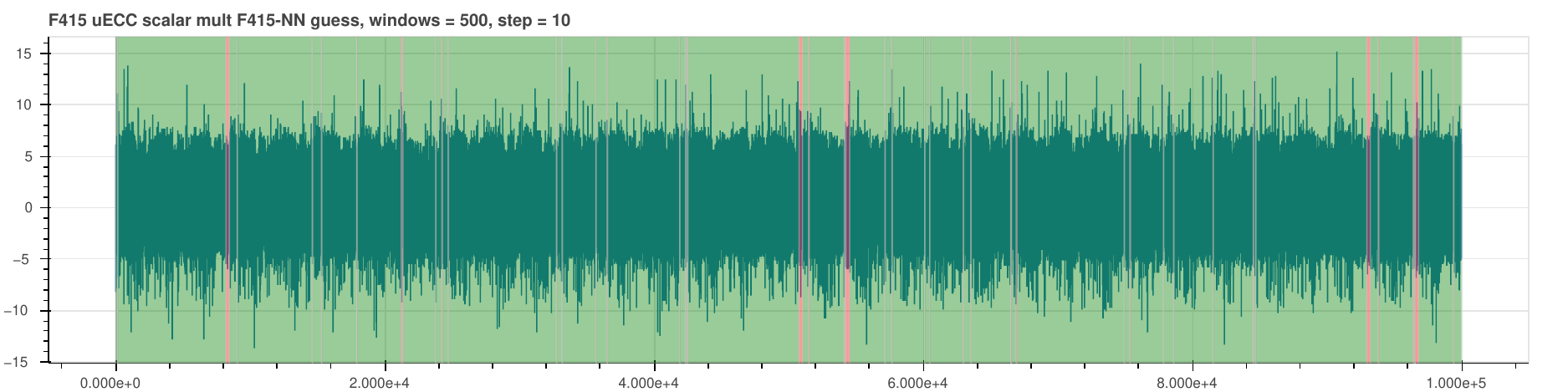}}
    \caption{Accuracy results for the initial portion of a real power consumption trace. In green the correct predictions, in red the failed ones.}
    \label{fig:accuracy_plot}
\end{figure}

Our network performed extremely well, as shown in Table \ref{tab:lstm_vs_cnn}, also on real-world traces, reaching an accuracy of around 97\% for each trace. 
In~\autoref{fig:accuracy_plot}, we plot the accuracy results for the initial portion of a real power consumption trace used during the testing phase.

From~\autoref{fig:accuracy_plot}, it is clear that the network was correct in most cases
(the green part of the plot). 

In the field of side-channel attack research, CNNs have been the predominant choice for analytical tasks. To demonstrate the superiority of the LSTM for this specific task, we performed an ablation study in which we removed the LSTM layer, thus converting the architecture into a CNN and using the same network parameters. In \ref{tab:lstm_vs_cnn}, we present the accuracy and loss results of our LSTM model when tested on real power consumption traces, and compare these with the results of a trained CNN model obtained by excluding the LSTM layer from the architecture mentioned in Section \ref{sec:sys}. This comparison validates our hypothesis that, given the similarity to the HAR problem, the LSTM is more suitable than the traditional CNN.

\begin{table}[!ht]
    \caption{Accuracy and loss of LSTM and CNN collected for testing real power consumption traces.}\label{tab:lstm_vs_cnn}
\centering
\scalebox{0.60}{
\resizebox{\columnwidth}{!}{
\begin{tabular}{r||l|l||l|l}
\emph{Trace} & \emph{LSTM Accuracy}& \emph{CNN Accuracy} & \emph{LSTM Loss} & \emph{CNN Loss} \\
\hline
\texttt{T1} & 0.9758 & 0.9437 & 0.0621 & 0.1763 \\
\texttt{T2} & 0.9766 & 0.9463 & 0.0603 & 0.1692 \\
\texttt{T3} & 0.9756 & 0.9499 & 0.0646 & 0.1746 \\
\texttt{T4} & 0.9728 & 0.9344 & 0.0702 & 0.2039 \\

\end{tabular}%
}
}
\end{table}

\subsection{End-To-End Attack}
An important issue encountered during the development of our system is the improper classification of successive operations of the same type (false positive). This issue arises from the neural network's lack of training to recognize the start and end of an operation. It has only been trained to distinguish between short- and long-term operations. As a result, when the model is faced with a sequence of \emph{Sub} operations, it is unable to discern which are short and which are long.

To overcome these limitations, we have adjusted our post-processing approach to operate on sequences rather than individual operations. We will refer to these sequences as \emph{clusters}, using them to identify the key bits. Specifically, we consider a trace to be valid if and only if all operations within it exhibit the same behavior (we define these as homogeneous clusters). This characteristic can be readily determined by comparing the cluster's length with the durations of short and long operations.

A significant disadvantage of this approach is the increased number of traces required because we are limited to analyzing traces that display homogeneous clusters. The decrease in the number of useful traces for any individual cluster is directly proportional to the cluster's length. Specifically, for a cluster of length $n$, the likelihood of obtaining a useful trace is $1/n$. Consequently, for $\Gamma$, the set of groups examined, the probability of finding a single trace where all clusters are homogeneous is $\prod_{\gamma \in \Gamma} 1/len(\gamma)$.

In order to quantify the increase in the number of traces needed for a successful attack, we recall ~\autoref{tab:key_op01}. We can clearly observe that operation 18 is part of a cluster of three \emph{Subs}, those at positions 17, 18, and 19. Instead, operation 26 is part of a cluster of 1. Then, operations 55, 88, 121, 154 and so on (each is 33 operations away),  are always in a cluster of 1, while operations 59, 92, 125, 158, etc. (each is 33 operations away) are always in a cluster of two (the \emph{Sub} before the change of the bit and the \emph{Sub} afterwards). This means that the first bit depends on the correct identification of the 3\textsuperscript{rd} and 5\textsuperscript{th} clusters. Also, since the 4\textsuperscript{th} cluster is a cluster of 3 operations, while the 5\textsuperscript{th} is a cluster of one, the attacker has a chance of 1/4 to get a useful prediction for a given signature trace. Instead, subsequent collisions depend on a cluster of 1 (operations 55, 88, 121, 154, etc.\ldots) and a cluster of two (operations 59, 92, 125, 158, etc\ldots). So, in this case, the predictions have probability one-half to be useful.
We also remark that from a useful collision we are able to correctly identify the exact value of the 2\textsuperscript{nd} bit of the ephemeral key, while for the next bit a useful collision only provides equality with the previous bit value. Thus, to obtain the value of the 3\textsuperscript{rd} bit of the ephemeral key, the attacker needs to get a useful collision on the 2\textsuperscript{nd} bit and a useful collision on the 3\textsuperscript{rd} bit. Similarly, to reveal the 4\textsuperscript{th} bit, three consecutive useful collisions on the same signature are needed.
\autoref{tab:bits_proba} summarizes these results. In particular, we observe that to obtain one signature where the attacker is able to retrieve 5 bits, she needs, on average, 64 signatures. This in turn means that to obtain 100 such signatures in order to perform the lattice reduction $6,400$ executions, it should be needed. In our scenario, execution of $6,400$ takes almost 10 days using the crypto algorithm on the Teledyne Lecroix 3104z oscilloscope model. Albeit this is an increase in the number of traces that should be acquired, this number is far from unfeasible in the context of a side-channel attack meaning that the solution does not limit the exploitation capability of the designed system.

\begin{table}[!ht]
    \caption{Probability to obtain \emph{n} bits for each signature, and estimation on the number of signatures required to obtain them.}\label{tab:bits_proba}
\centering
\scalebox{0.6}{
\resizebox{\columnwidth}{!}{
\begin{tabular}{r|l|l}
\#\emph{Bits} & \emph{Probability} & \emph{Estimated sign. required}\\
\hline
1 & $\sfrac{1}{4}$ & 4 \\
2 & $\sfrac{1}{4} \cdot \sfrac{1}{2} = \sfrac{1}{8}$ & 8\\
3 & $\sfrac{1}{4} \cdot \sfrac{1}{2} \cdot \sfrac{1}{2} = \sfrac{1}{16}$ & 16\\
4 & $\sfrac{1}{4} \cdot \sfrac{1}{2} \cdot \sfrac{1}{2} \cdot \sfrac{1}{2} = \sfrac{1}{32}$ & 32\\
5 & $\sfrac{1}{4} \cdot \sfrac{1}{2} \cdot \sfrac{1}{2} \cdot \sfrac{1}{2} \cdot \sfrac{1}{2} = \sfrac{1}{64}$ & 64\\

\end{tabular}%
}
}
\end{table}

\section{Countermeasures}
\label{sec:countermeasures}

In this Section, we discuss a range of countermeasures suitable for various types of implementation. These countermeasures vary in their requirements for random generation and differ in efficiency levels.

\subsection{Incomplete Countermeasures}
\paragraph{Real effectiveness of the~\cite{luo2018effective} countermeasure.}
If the implementation can afford using one additional register, it is possible to adopt the countermeasure suggested by Luo~\textit{et al.}~\cite{luo2018effective}. 
Since the weakness arises from the subtraction computation in XYCZ-ADDC, and since the result is squared,
it does not matter if it is $X_0-X_1$ or $X_1-X_0$. So, the authors suggest saving the subtraction result performed during XYCZ-ADD, and reuse it in XYCZ-ADDC for the next bit. 
We remark that this countermeasure is very efficient and also saves one subtraction. However, XYCZ-ADDC subtraction is not the only collision exploitable during execution. For example, we observe that the order of use of $X_1$ and $X_2$ in iteration $i$ in XYCZ-ADDC depends on the value of the key bits $k_{b_{i-1}}$ and $k_{b_i}$. So, for example, if $k_{b_{i-1}}$ equals $k_{b_i}$, the $X$ outputs of XYCZ-ADD will be used in the same order as they have been computed; otherwise, their order will be reversed. Using collisions, an attacker can detect if the same value is used during the first or second computation (that is, when computing $B$ or $C$), and infer the values of the key bits. 

\paragraph{Modular Reduction.}
A classic countermeasure that can be applied to the implementation is to always perform modular reductions, regardless of whether the computation overflows or underflows. This approach was previously suggested in \cite{ryan2019return} and is probably the easiest solution to this problem. However, this countermeasure would not prevent a side-channel attacker from detecting that the same value has been computed twice (that is, a collision on the values) through correlation techniques on the power or EM traces. This is the kind of weakness exploited in~\cite{luo2018effective} and, as observed in Section~\ref{sec:attackz}, it would again make it possible for an attacker to obtain sensitive information. 

\subsection{Effective Countermeasures}
\paragraph{Masking Technique.}
A more robust approach to counteract the attack is to mask the values (that is, the coordinates) manipulated during the execution of the Montgomery ladder. This method is more effective, but also less efficient. To avoid collisions, masks must be changed for each iteration of the Montgomery ladder. Otherwise, an attacker could detect the occurrence of a collision regardless of the presence of the masks. 

\paragraph{Coordinates Re-Randomization.}
Another solution to this problem is to repeat randomization of the coordinates after each iteration of the Montgomery loop. This approach involves generating a new \emph{z} for each iteration and reprojecting the curve to the new coordinates. However, it could be more cost-effective than a fully masked implementation and may also prevent collision attacks. The process of re-randomizing the coordinates essentially ensures that there is no correlation between the previous and current computations, and thus eliminates the possibility of a collision attack. 

\section{Conclusions}
Our study rigorously evaluated the implementation of ECDSA on a commercial standard device using the micro-ecc library, focusing on the secp160r1 curve. We identified vulnerabilities, particularly in non-constant-time execution, due to selective modular reductions. By improving existing collision attacks and applying them to real-world hardware, we exposed more flaws. Using simulations and an LSTM neural network, we detected operational patterns indicating modular reductions, enabling us to derive ephemeral key bits and recover the signing key through lattice attacks. Testing on a real STM32F415RG device with the ChipWhisperer confirmed our method's high accuracy and effectiveness in revealing significant security risks in prevalent cryptographic systems.


\end{document}